\documentclass[aps,pra,twocolumn,a4paper,showpacs,superscriptaddress,longbibliography]{revtex4-1}

\usepackage{amsmath,amssymb,bm,graphicx}
\usepackage{xcolor}
\usepackage[colorlinks=true,urlcolor=blue,linkcolor=blue,citecolor=blue,bookmarks=false]{hyperref}

\begin{document}

\title{Canted antiferromagnetism in phase-pure CuMnSb}

\author{A.~Regnat}
\email{alexander.regnat@frm2.tum.de}
\affiliation{Physik-Department, Technische Universit\"{a}t M\"{u}nchen, D-85748 Garching, Germany}

\author{A.~Bauer}
\email{andreas.bauer@frm2.tum.de}
\affiliation{Physik-Department, Technische Universit\"{a}t M\"{u}nchen, D-85748 Garching, Germany}

\author{A.~Senyshyn}
\affiliation{Heinz Maier-Leibnitz Zentrum (MLZ), Technische Universit\"{a}t M\"{u}nchen, D-85748 Garching, Germany}

\author{M.~Meven} 
\affiliation{Institut f\"{u}r Kristallographie, RWTH Aachen, D-52056 Aachen, Germany}
\affiliation{Forschungszentrum J\"{u}lich GmbH, J\"{u}lich Centre for Neutron Science~(JCNS) at Heinz Maier-Leibnitz Zentrum (MLZ), 85748 Garching, Germany}

\author{K. Hradil}
\affiliation{Center of X-ray Technology, Vienna University of Technology, A-1060 Vienna, Austria}

\author{P.~Jorba}
\affiliation{Physik-Department, Technische Universit\"{a}t M\"{u}nchen, D-85748 Garching, Germany}

\author{K.~Nemkovski}
\affiliation{Forschungszentrum J\"{u}lich GmbH, J\"{u}lich Centre for Neutron Science~(JCNS) at Heinz Maier-Leibnitz Zentrum (MLZ), 85748 Garching, Germany}

\author{B.~Pedersen} 
\affiliation{Heinz Maier-Leibnitz Zentrum (MLZ), Technische Universit\"{a}t M\"{u}nchen, D-85748 Garching, Germany}

\author{R.~Georgii}
\affiliation{Physik-Department, Technische Universit\"{a}t M\"{u}nchen, D-85748 Garching, Germany}
\affiliation{Heinz Maier-Leibnitz Zentrum (MLZ), Technische Universit\"{a}t M\"{u}nchen, D-85748 Garching, Germany}

\author{S.~Gottlieb-Sch\"{o}nmeyer}
\affiliation{Physik-Department, Technische Universit\"{a}t M\"{u}nchen, D-85748 Garching, Germany}

\author{C.~Pfleiderer}
\affiliation{Physik-Department, Technische Universit\"{a}t M\"{u}nchen, D-85748 Garching, Germany}

\date{\today}

\begin{abstract}
We report the low-temperature properties of phase-pure single crystals of the half-Heusler compound CuMnSb grown by means of optical float-zoning. The magnetization, specific heat, electrical resistivity, and Hall effect of our single crystals exhibit an antiferromagnetic transition at $T_{\mathrm{N}} = 55$~K and a second anomaly at a temperature $T^{*} \approx 34$~K. Powder and single-crystal neutron diffraction establish an ordered magnetic moment of $(3.9\pm0.1)~\mu_{\mathrm{B}}/\mathrm{f.u.}$, consistent with the effective moment inferred from the Curie-Weiss dependence of the susceptibility. Below $T_{\mathrm{N}}$, the Mn sublattice displays commensurate type-II antiferromagnetic order with propagation vectors and magnetic moments along $\langle111\rangle$ (magnetic space group $R[I]3c$). Surprisingly, below $T^{*}$, the moments tilt away from $\langle111\rangle$ by a finite angle $\delta \approx 11^{\circ}$, forming a canted antiferromagnetic structure without uniform magnetization consistent with magnetic space group $C[B]c$. Our results establish that type-II antiferromagnetism is not the zero-temperature magnetic ground state of CuMnSb as may be expected of the face-centered cubic Mn sublattice.
\end{abstract}

\maketitle


\section{Motivation}
The properties of magnetic atoms arranged in a face-centered cubic (fcc) lattice have been a cornerstone for the development of an understanding of antiferromagnetic order in solids. Anticipated theoretically on the basis of measurements of the magnetic susceptibility~\cite{1932:Neel:AnnPhysParis, 1938:Bitter:PhysRev, 1941:vanVleck:JChemPhys, 1950:Anderson:PhysRev, 1971:Neel:Science}, neutron scattering in the transition metal oxides MnO, NiO, and CoO~\cite{1949:Shull:PhysRev, 1951:Shull:PhysRev} marked the starting point of the microscopic discovery of antiferromagnetism~\cite{1954:Lidiard:RepProgPhys, 1955:Nagamiya:AdvPhys}. Crystallizing in a NaCl structure, the transition metal ions in these systems form a fcc sublattice, supporting $\langle111\rangle$ magnetic order, now also referred to as fcc type-II antiferromagnetism. With the development of the notion of super-exchange interactions in transition metal oxides~\cite{1934:Kramers:Physica, 1959:Anderson:PhysRev, 1959:Kanamori:JPhysChemSolids, 2001:Pask:PhysRevB}, a well-founded qualitative and quantitative account of the underlying interactions driving antiferromagnetism on a more general level was initiated. In turn, materials featuring fcc sublattices of the magnetic atoms offer an important point of reference for the identification of new facets of antiferromagnetic ordering phenomena.

A materials class in which the physical properties of fcc sublattices are at the heart of an exceptionally wide range of properties are Heusler compounds, cf.\ Ref.~\onlinecite{2011:Graf:ProgSolidStateCh} for a comprehensive review. Two major subclasses may be distinguished. Full-Heusler compounds, $X_{2}YZ$, crystallize in the L2$_{1}$ structure (space group $Fm\bar{3}m$), comprising four fcc sublattices at $(0,0,0)$, $(1/4,1/4,1/4)$, $(1/2,1/2,1/2)$, and $(3/4,3/4,3/4)$. In comparison, half-Heusler compounds, $XYZ$, crystallize in the non-centrosymmetric C1$_{\textrm{b}}$ structure (space group $F\bar{4}3m$), where one of the fcc sublattices is vacant. Quaternary as well as so-called inverse Heusler compounds also belong to space group $F\bar{4}3m$. The constituent elements $X$, $Y$, and $Z$ may be selected from large parts of the periodic table, which permits to realize a wide range of physical properties reaching from metallic to insulating behavior and from diamagnetism to ferromagnetism. Further examples include superconductors, thermoelectrics, heavy-fermion compounds, and shape memory alloys, even topological insulators are predicted. Since all of these properties may be achieved within the same crystal structure, all-Heusler-devices are promising goals for future research.

Antiferromagnetism in Heusler compounds is rather rare, in particular when compared to prevalent ferromagnetism. Known examples are typically based on $5f$ or $6f$ elements~\cite{1991:Adroja:JMagnMagnMater, 1996:Seaman:PhysRevB, 2005:Gofryk:SolidStateCommun, 2005:Gofryk:PhysRevB, 2008:Casper:ZanorgallgChem}, in which the magnetism is carried by localized $f$ electrons, or heavy $4d$ or $5d$ transition metals~\cite{1968:Webster:JApplPhys, 1971:Hames:JApplPhys, 1972:Masumoto:JPhysSocJpn, 1975:Campbell:JPhysFMetPhys, 1986:Helmholdt:JLessCommonMet}. 

Among half-Heusler compounds based on $3d$ transition metals, CuMnSb may be the only example exhibiting antiferromagnetism~\cite{1968:Endo:JPhysSocJpn}. At low temperatures, Forster \textit{et al.} reported antiferromagnetic order in polycrystalline samples with large ordered moments of $(3.9\pm0.1)~\mu_{\mathrm{B}}/\mathrm{f.u.}$, in which ferromagnetic $\{111\}$ planes of $\langle111\rangle$-oriented manganese moments couple antiferromagnetically with their neighboring planes~\cite{1968:Forster:JPhysChemSolids}. Thus, the antiferromagnetism in CuMnSb represents precisely the type-II form observed in the transition metal oxides. In view of the large ordered moments on the manganese site this behavior suggests local-moment magnetism. However, as the C1$_b$ crystal structure lacks inversion symmetry, additional weak spin--orbit interactions may be present. Further, CuMnSb is metallic.

Recently, ab initio calculations by M\'{a}ca \textit{et al.} based on a Heisenberg model demonstrated that type-II antiferromagnetism may in fact not be the ground state of perfectly ordered CuMnSb~\cite{2016:Maca:PhysRevB}. Instead, antiferromagnetic states with moments oriented along $\langle100\rangle$, a tetragonal arrangement with alternating double layers of opposite spins along the $\langle210\rangle$ directions, or even more complex spin states are expected. Point defects, however, may stabilize the antiferromagnetic structure observed experimentally. The authors determine that in particular Mn antisites on the Cu lattice and Mn interstitials favor ordered moments along $\langle111\rangle$ already at concentrations of a few percent, i.e., values that are, for instance, consistent with the relatively large residual resistivity of ${\sim}50~\mu\Omega$cm reported in the literature~\cite{1982:Schreiner:SolidStateCommun, 1989:Otto:JPhysCondensMatter, 1995:Kirillova:PhysStatusSolidiB, 2006:Boeuf:PhysRevB, 2012:Maca:JMagnMagnMater}. Such pronounced sensitivity of the physical properties on defects and disorder is a quite common phenomenon in Heusler compounds, resulting in modified or even drastically altered physical properties as compared to the expectations for the perfectly ordered host material~\cite{1999:Orgassa:PhysRevB, 2004:Miura:PhysRevB, 2004:Picozzi:PhysRevB, 2011:Graf:ProgSolidStateCh}. Unraveling this intimate connection, in turn, seems to be essential for both technological applications and the fundamental understanding of the underlying physics.

In this context, the nature and origin of the antiferromagnetism in CuMnSb is also puzzling in so far, as the low-temperature properties combine characteristics that are believed to be hallmarks of either local-moment or itinerant magnetism~\cite{2003:Pfleiderer:PhysicaB, 2006:Boeuf:PhysRevB}. Notably, large ordered moments and the commensurate magnetic structure are contrasted by metallic electrical resistivity, a relatively low transition temperature, and a peculiar stability of the magnetic order in magnetic fields~\cite{2004:Doerr:PhysicaB}. In addition, distinctly different values of the N\'{e}el temperature $T_{\textrm{N}}$ (50~K, 55~K, 62~K), the fluctuating moment $m_{\textrm{eff}}$ ($6.3~\mu_{\mathrm{B}}/\mathrm{f.u.}$, $5.6~\mu_{\textrm{B}}/\mathrm{f.u.}$, $5.2~\mu_{\mathrm{B}}/\mathrm{f.u.}$), and the Curie-Weiss temperature $\mathit{\Theta}_{\mathrm{CW}}$ ($-250$~K, $-160$~K, $-120$~K) were reported by B{\oe}uf \textit{et al.}~\cite{2006:Boeuf:PhysRevB}, Endo~\cite{1970:Endo:JPhysSocJpn}, and Helmholdt \textit{et al.}~\cite{1984:Helmholdt:JMagnMagnMater}. Presumably, these discrepancies are also related to structural disorder or the presence of a magnetic impurity phase~\cite{1970:Endo:JPhysSocJpn}. 

Speculations about a half-metallic character, in analogy to isostructural NiMnSb~\cite{1983:deGroot:PhysRevLett, 1995:vanLeuken:PhysRevLett}, were clarified by electronic structure calculations that reproduced the antiferromagnetically ordered moments of $4~\mu_{\mathrm{B}}/\mathrm{f.u.}$ and identified CuMnSb as a compensated semi-metallic compound~\cite{2005:Jeong:PhysRevB}. According to Jeong \textit{et al.}, the antiferromagnetic phase may be pictured heuristically as self-doped Cu$^{1+}$Mn$^{2+}$Sb$^{3-}$ and is characterized by a small, semi-metallic density of states at the Fermi level with a finite minority band occupation. A large electron mass enhancement is predicted due to spin fluctuations while the hole masses are expected to stay normal~\cite{2005:Jeong:PhysRevB, 2007:Podgornykh:JMagnMagnMater}.

These spin fluctuations may also be associated with the comparatively low transition temperature in a compound that is in close chemical and structural proximity to high-temperature ferromagnets, such as NiMnSb, and promising candidates for antiferromagnetic spintronics, such as CuMnAs~\cite{2016:Wadley:Science}. Thus, understanding the nature of magnetism in CuMnSb may not only contribute to the design of tailored antiferromagnetic Heusler compounds, but also help to capture the mutual interplay of local-moment and itinerant magnetism and provide fresh input to the wide field of research in which spin fluctuations play an important role. Despite the multitude of unresolved issues, in particular concerning the origin of the antiferromagnetism, and the apparent sensitivity of the physical properties on sample quality, however, only polycrystalline specimens were studied in the literature so far. Open questions thereby concern whether the puzzling combination of properties persists in phase-pure single crystals and whether a commensurate $\langle111\rangle$-oriented antiferromagnetic structure is the magnetic ground state. 

In our paper, we report single-crystal growth of CuMnSb by means of the optical floating-zone technique. Careful X-ray powder diffraction allows us to identify the tetragonal ferrimagnet Mn$_{2}$Sb as impurity phase in polycrystalline specimens. Its formation may be suppressed by a small excess of antimony in the initial starting composition as reported in Ref.~\onlinecite{1970:Endo:JPhysSocJpn}. Measurements on our phase-pure single crystals permit to resolve several ambiguities arising from earlier reports. We find fluctuating moments of nearly 4~$\mu_{\mathrm{B}}/\mathrm{f.u.}$ inferred from the Curie-Weiss-like temperature dependence of the magnetization, ordered moments of the same size inferred from neutron scattering, and a corresponding contribution to the entropy of $R\,\ln9$, consistently implying a local-moment character of the magnetism in CuMnSb.
 
As an unexpected new aspect magnetization, specific heat, and transport data further identify a change in the electronic or magnetic structure at a temperature $T^{*} \approx 34$~K, well below the onset of antiferromagnetic order at the N\'{e}el temperature $T_{\mathrm{N}} = 55$~K. Using powder and single-crystal neutron diffraction, we are able to unambiguously attribute this observation to a canting of the commensurate antiferromagnetic state without uniform magnetic moment. The magnetic space group thereby changes from $R[I]3c$ for $T^{*} < T < T_{\mathrm{N}}$ to $C[B]c$ for $T < T^{*}$. Thus, the canted antiferromagnetism below $T^{*}$ is consistent with the predictions by M\'{a}ca \textit{et al.}~\cite{2016:Maca:PhysRevB}.

Our paper is organized as follows. In Sec.~\ref{Methods}, we describe the sample preparation and the experimental methods used in this study. In Sec.~\ref{Results}, we address the low-temperature properties of CuMnSb. We begin with the magnetization and a discussion of the magnetic anisotropy, before we continue with data on the specific heat and entropy. Next, we turn to the transport properties, namely electrical resistivity, thermal conductivity, Seebeck coefficient, and Hall effect. As a central aspect, we resolve the magnetic structure of phase-pure CuMnSb and its change as a function of temperature by means of powder and single-crystal neutron diffraction data. Finally, we summarize our findings in Sec.~\ref{Conclusions}.


\section{Experimental methods}
\label{Methods}

\subsection{Single-crystal growth and characterization}

\begin{figure}
\includegraphics[width=1.0\linewidth]{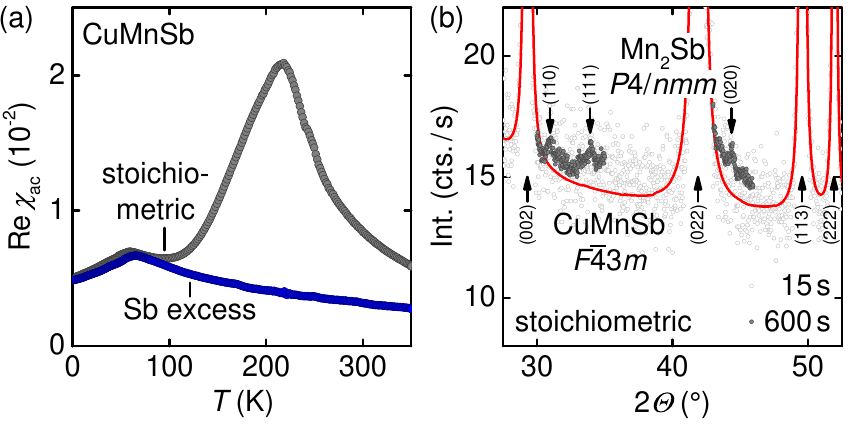}
\caption{Signatures of a Mn$_{2}$Sb impurity phase in CuMnSb prepared from stoichiometric weight ratio of the starting elements. (a)~Real part of the susceptibility as a function of temperature for two float-zoned samples prepared from stoichiometric initial weight (gray curve) and with slight antimony excess (blue curve), respectively. The broad maximum around 200~K indicates the presence of a magnetically ordering impurity phase. (b)~High-resolution X-ray powder diffractogram obtained through long counting times. Measuring where no Bragg peaks are expected in the space group $F\bar{4}3m$ of CuMnSb, we identify the impurity phase as tetragonal Mn$_{2}$Sb.}
\label{figure01}
\end{figure}

The objective of our study was the preparation of phase-pure high-quality single crystals of CuMnSb. In a seminal study, Endo reported the observation of a broad maximum in the susceptibility around 200~K in arc-melted CuMnSb samples prepared from a stoichiometric weight ratio of the starting elements~\cite{1970:Endo:JPhysSocJpn}. This maximum was attributed to the presence of a Mn$_{2}$Sb impurity phase. In pure form, Mn$_{2}$Sb crystallizes in the tetragonal space group $P4/nmm$ and displays ferrimagnetic order with a Curie temperature $T_{\mathrm{C}} = 550$~K, followed by a change of the easy direction from the $c$-axis at high temperatures to the basal plane around 240~K~\cite{1957:Wilkinson:JPhysChemSolids}. While the metallurgical details of this putative Mn$_{2}$Sb phase could not clarified, it was noted that the formation of the impurity phase could be suppressed empirically by a small antimony excess in the starting elements, where the lattice constant of CuMnSb was found to be unchanged.

For our study single crystals of CuMnSb were grown by means of optical float-zoning using an ultra-high vacuum compatible preparation chain~\cite{2016:Bauer:RevSciInstrum2}. The preparation started from high-purity elements (6N copper, precast 4N manganese, and 6N antimony). Polycrystalline feed rods were cast by means of an inductively heated rod casting furnace~\cite{2016:Bauer:RevSciInstrum}. The feed rods were optically float-zoned at a rate of 5~mm/h while counter-rotating at 6~rpm in a high-purity argon atmosphere of 2.5~bar~\cite{2011:Neubauer:RevSciInstrum}. No evaporation losses were observed during sample preparation, presumably due to the relatively low melting point of CuMnSb of ${\sim}800~^{\circ}\mathrm{C}$.

Figure~\ref{figure01}(a) shows that the ac susceptibility of float-zoned CuMnSb as prepared from a stoichiometric ratio of the starting elements (gray symbols). A pronounced maximum around 200~K is observed, characteristic of the impurity phase. Standard X-ray diffraction with a Siemens D5000 diffractometer using copper $K_{\alpha1}$ radiation, cf.\ open symbols and solid line in Fig.~\ref{figure01}(b), suggested at first sight phase-pure CuMnSb with space group $F\bar{4}3m$ with volume fractions of impurity phases below the detection limit. However, for long exposure times at diffraction angles where CuMnSb possesses no intensity maxima (solid symbols) tiny additional peaks could be observed. The position of these peaks were characteristic of tetragonal Mn$_{2}$Sb suggesting a volume fraction well below 1\%. This finding highlights the ac susceptibility as an exceptionally sensitive probe for the detection of magnetically ordering impurity phases in systems with a very low intrinsic susceptibility, such as CuMnSb.  

Further, as shown in Fig.~\ref{figure01}(a), for a preparation process with a starting weight ratio of Cu : Mn : Sb of 1 : 1 : 1.035 the susceptibility no longer displays the anomalous contribution at high temperature (blue symbols). Comparison with the susceptibility data of the sample prepared from the stoichiometric starting composition suggests as an upper boundary of the volume fraction of Mn$_{2}$Sb less than 0.01\%. Hence, consistent with the results of Endo~\cite{1970:Endo:JPhysSocJpn}, we find that the formation of Mn$_{2}$Sb may be suppressed by a small antimony excess in the starting composition, permitting the preparation of phase-pure CuMnSb. All results presented in the following were obtained on such phase-pure samples.

\begin{figure}
\includegraphics[width=1.0\linewidth]{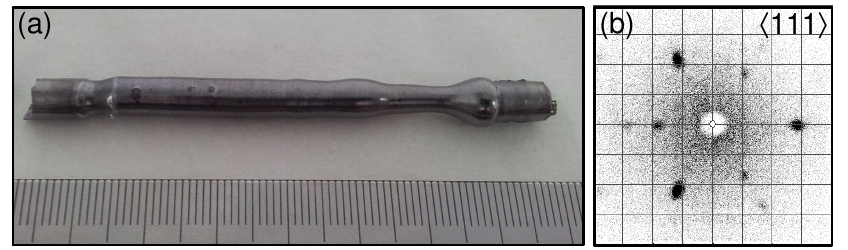}
\caption{Key characteristics of the single crystal grown for this study. (a)~Optically float-zoned single crystal of CuMnSb. The scale on the bottom is given in millimeter. Growth direction was from right to left. (b)~Threefold X-ray Laue pattern of a cubic $\langle111\rangle$ axis.}
\label{figure02}
\end{figure}

Shown in Fig.~\ref{figure02}(a) is a photograph of the float-zoned ingot. Using X-ray Laue diffraction, the final 15~mm of the ingot were identified as a single crystal across the entire cross-section, cf.\ Fig.~\ref{figure02}(b). We note that copper fluorescence limited the quality of the Laue pictures somewhat. From the end of the single crystal, a disc of 1~mm thickness was cut with a $\langle110\rangle$ direction along its axis. Starting with this disc, a bar of $6\times1\times1~\mathrm{mm}^{3}$ and a platelet of $6\times1\times0.2~\mathrm{mm}^{3}$ were prepared, both with their longest edge parallel to $\langle100\rangle$ and their shorter edges parallel to $\langle110\rangle$. Measurements of the ac susceptibility, magnetization, specific heat, and thermal transport were carried out on the bar-shaped sample. The electrical transport properties were studied on the platelet. In addition, a cuboid of $3\times3\times2~\mathrm{mm}^{3}$ with two faces oriented along $\langle100\rangle$ and four along $\langle110\rangle$ was prepared from the bottom of the single-crystal ingot. This sample was used in our neutron diffraction studies at HEiDi. The remaining tilted cylinder with a diameter and height of 6~mm and 11~mm, respectively, was used for neutron diffraction at RESI, MIRA, and DNS. All data presented in the following are shown as recorded, without correcting the effects of demagnetizing fields, since the absolute value of the magnetization in CuMnSb is tiny.

\begin{figure}
\includegraphics[width=1.0\linewidth]{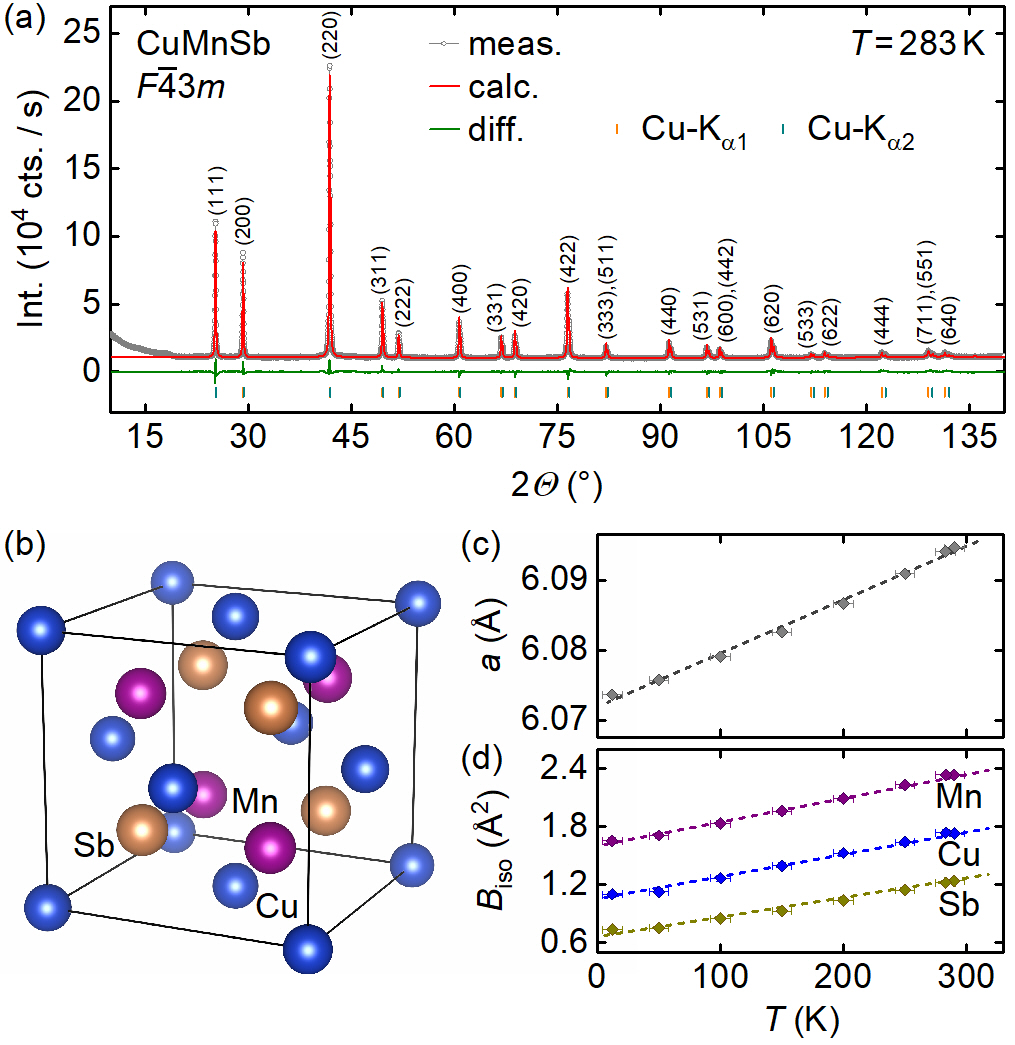}
\caption{Salient properties observed in X-ray powder diffraction. (a)~Powder diffractogram of float-zoned material. Measured data (open symbols) and a Rietveld refinement based on the cubic half-Heusler space group $F\bar{4}3m$ (solid red line) are in excellent agreement. (b)~Schematic depiction of the crystal structure. (c)~Lattice constant $a$ as a function of temperature. (d)~Isotropic displacement parameters $B_{\mathrm{iso}}$ as a function of temperature. The dashed lines are linear guides to the eye.}
\label{figure03}
\end{figure}

X-ray powder diffraction at different temperatures was carried out on a Bruker D8 Advance with a closed-cycle cryostat using copper $K_{\alpha1}$ and $K_{\alpha2}$ radiation. Shown in Fig.~\ref{figure03}(a) are typical data collected near room temperature. A Rietveld refinement (solid red line) is in excellent agreement with phase-pure CuMnSb with the half-Heusler space group $F\bar{4}3m$, cf.\ Fig.~\ref{figure03}(b). Down to the lowest temperatures studied, no hints suggesting a structural phase transition were observed. As shown in Fig.~\ref{figure03}(c), the cubic lattice constant $a$ monotonically decreases from $6.095~\textrm{\AA}$ to $6.074~\textrm{\AA}$, consistent with values reported in the literature~\cite{1952:Nowotny:MonatshChem, 1952:Castelliz:MonatshChem}. The isotropic displacement parameters for the three elements, see Fig.~\ref{figure02}(d), also decrease linearly as a function of decreasing temperature. Their relatively large absolute values suggests considerable structural disorder as discussed in the following.

Simultaneous refinement of the X-ray and neutron powder data, cf.\ Sec.~\ref{Powder}, permitted to obtain a quantitative estimate of the dominant type of defects. In our assessment we considered antisite disorder in the form of mixed occupancies at two atomic sites at a time. The data was refined with respect to the degree of mixing as free parameter. As the main result we find 1.5\% antisite disorder of the Cu and Mn atoms, in excellent agreement with reports of 1.6\% of Cu/Mn antisite disorder in samples investigated by M\'{a}ca \textit{et al.}~\cite{2016:Maca:PhysRevB}. Such defect concentrations may seem high for intermetallic compounds, but in fact are rather low for half-Heusler compounds.


\subsection{Low-temperature bulk and transport properties}

The ac susceptibility was measured in a 9~T Quantum Design physical properties measurement system (PPMS) at an excitation frequency of 911~Hz and an excitation amplitude of 1~mT. The magnetization was determined in a 9~T Oxford Instruments vibrating sample magnetometer, using an excitation amplitude of about 1~mm at a frequency of 62.35~Hz. The specific heat was measured in a 14~T PPMS using a small-pulse method, where the heat pulses had a typical size of 0.5\% of the current sample temperature.

For measurements of the thermal conductivity and the Seebeck coefficient, flattened silver wires of 0.25~mm diameter were glued onto the bar-shaped sample in a four-terminal configuration (heater, hot thermometer, cold thermometer, cold bath) using silver epoxy~\cite{2013:Gangl:Master}. Utilizing the thermal transport option of the 14~T PPMS, heat pulses of ${\sim}3\%$ of the current temperature were applied while continuously sweeping the temperature from 300~K to 2~K at a rate of 0.5~K/min.

For measurements of the electrical resistivity and the Hall effect, gold wires of $25~\mu$m diameter were spot-welded onto the platelet sample in a six-terminal configuration. The measurements were performed in a 14~T Oxford Instruments magnet system using a standard low-current digital lock-in technique at an excitation frequency of 22.08~Hz. The electrical resistivity, $\rho_{xx}$, and the Hall resistivity, $\rho_{xy}$, were inferred from the longitudinal and transverse voltage pick-up following symmetrization and anti-symmetrization, respectively~\cite{2013:Ritz:PhysRevB}. All geometry factors were determined from digital photographs of the samples and sample contacts recorded with an optical microscope.


\subsection{Neutron diffraction}

All neutron scattering data presented in this paper were collected at the Heinz Maier-Leibnitz Zentrum~(MLZ). Powder diffraction was performed on the high-resolution powder diffractometer SPODI~\cite{2012:Hoelzel:NuclInstrumMethA, 2015:Hoelzel:JLSFR} in Debye-Scherrer geometry at an incident neutron wavelength of $1.548~\textrm{\AA}$. The sample was prepared from float-zoned material by means of a cryomill and filled into a thin-walled vanadium cylinder. The diffraction data were corrected for geometrical aberrations and the curvature of the Debye-Scherrer rings.

Single-crystal diffraction was carried out on the 4-circle single-crystal diffractometers RESI~\cite{2015:Pedersen:JLSFR} and HEiDi~\cite{2015:Meven:JLSFR}. On RESI thermal neutrons with a wavelength of $1.041~\textrm{\AA}$ were used. A MAR345 image plate detector allowed for fast scans of reciprocal space planes, where the scattering intensities were integrated by means of the EVAL-14 method~\cite{2003:Duisenberg:JApplCrystallogr}. The temperature dependence of selected Bragg peaks was studied using a counting tube. On HEiDi hot neutrons with a wavelength of $0.794~\textrm{\AA}$ were used. All data were recorded with a counting tube and scattering intensities were integrated using the Lehmann-Larsen algorithm~\cite{1977:Will:JMagnMagnMater}.

The multi-purpose instrument MIRA~\cite{2015:Georgii:JLSFR, 2018:Georgii:NuclInstrumMethodsPhysResA} was used as a triple-axis spectrometer at an incident neutron wavevector of $1.396~\textrm{\AA}^{-1}$. We show data inferred from energy scans at the $\frac{1}{2}(111)$ position in reciprocal space. On the diffuse neutron scattering spectrometer with polarization analysis DNS~\cite{2001:Schweika:PhysicaB, 2015:Su:JLSFR}, we carried out diffraction experiments at an incident neutron wavelength of $4.2~\textrm{\AA}$, and the neutron polarization parallel to the momentum transfer, referred to as $x$-polarization, for the middle part of the detector bank. Additionally, in these diffraction experiments the clear separation in reciprocal space as well as the distinct evolution as a function of temperature allowed to distinguish between nuclear (non-spin-flip) and magnetic (spin-flip) reflections. Intensity maxima stemming from a $\lambda/2$ contamination of the incoming neutron beam were manually removed from the data. 

In all neutron scattering experiments, low temperatures were provided by top-loading closed-cycle cryostats. When not stated otherwise, the samples were glued on top of bespoke sample holders made of aluminum using a small amount of GE varnish. Data collected at SPODI, RESI, and HEiDi were analyzed by means of Rietveld and least-square refinements, respectively, using the software packages FullProf~\cite{1993:RodriguezCarvajal:PhysicaB} and Jana2006~\cite{2014:Petricek:ZKristallogr}.
 

\section{Experimental results}
\label{Results}

\subsection{Magnetization and specific heat}
\label{Magnetization}

\begin{figure}
\includegraphics[width=1.0\linewidth]{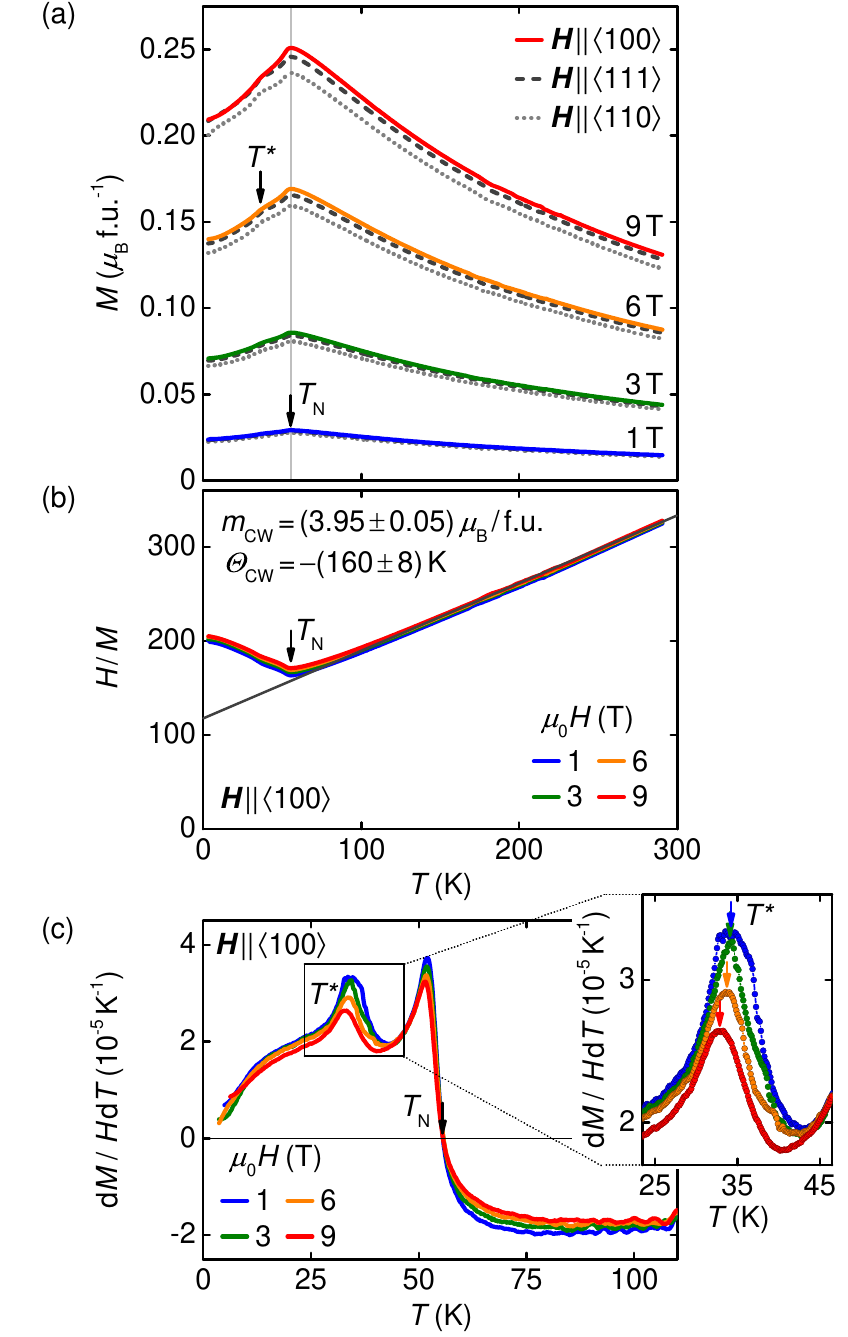}
\caption{Typical magnetization and susceptibility data. (a)~Temperature dependence of the magnetization for typical magnetic fields applied along the major crystallographic axes. A kink marks the N\'{e}el temperature $T_{\mathrm{N}}$. A cusp below $T_{\mathrm{N}}$ is referred to as $T^{*}$. (b)~Inverse normalized magnetization, $H/M$, as a function of temperature. The solid gray line indicates Curie-Weiss-like behavior at high temperatures. (c)~Temperature derivative of the normalized magnetization, $\mathrm{d}M/H\mathrm{d}T$, as a function of temperature illustrating how $T^{*}$ and $T_{\mathrm{N}}$ are inferred. Inset: Close-up view of the regime around $T^{*}$.}
\label{figure04}
\end{figure}

We begin the presentation of our experimental results with the magnetization shown in Fig.~\ref{figure04}(a). As a function of decreasing temperature, the magnetization increases up to a maximum that may be attributed to the onset of antiferromagnetic order at the N\'{e}el temperature $T_{\mathrm{N}}$, consistent with neutron diffraction presented below. Increasing the applied magnetic field increases the absolute value of the magnetization but leaves the shape of the curve unchanged. In particular, $T_{\mathrm{N}}$ is unchanged up to 9~T. A maximum value of the magnetization of 0.25~$\mu_{\mathrm{B}}/\mathrm{f.u.}$ at 9~T corresponds to $1/16$ of the ordered moment reported from neutron scattering~\cite{1968:Forster:JPhysChemSolids}. For different magnetic field directions, namely field parallel to $\langle100\rangle$ (solid line), $\langle110\rangle$ (dotted line), and $\langle111\rangle$ (dashed line), we observe essentially the same behavior, characteristic of isotropic magnetic properties. The absolute value of the magnetization varies by only ${\sim}5\%$ with $\langle100\rangle$ being magnetically softest and $\langle110\rangle$ being hardest.

The lack of field dependence of $T_{\mathrm{N}}$ and the small absolute value of the magnetization were previously discussed in view of an itinerant character of the magnetism in CuMnSb. In the context of the weak anisotropy of the system, however, we rather consider it as the hallmark of strong isotropic exchange interactions. Consequently, we expect that the magnetic moments smoothly rotate towards the field direction as a function of increasing magnetic field with a saturation field in excess of 100~T. This assumption is also corroborated by Ref.~\onlinecite{2004:Doerr:PhysicaB}, reporting an essentially linear increase of the magnetization as a function of fields up to 50~T.

In addition to the maximum at $T_{\mathrm{N}}$, we observe a small cusp at $T^{*} < T_{\mathrm{N}}$ for all field values and directions studied. Such an anomaly was not reported in earlier studies. As will be established below by means of neutron scattering, this signature may be attributed to an antiferromagnetic spin canting without emergence of a uniform magnetic moment. In previous studies, this delicate rearrangement of the magnetic structure, in fact representing the magnetic ground state of CuMnSb, presumably may have been suppressed by an abundance of point defects as suggested in a recent ab initio study~\cite{2016:Maca:PhysRevB}.

Shown in Fig.~\ref{figure04}(b) is the inverse normalized magnetization, $H/M$, as a function of temperature. Data for different magnetic field values collapse on a single curve, illustrating the essentially linear field dependence of the magnetization in the field range studied. Consequently, $H/M$ provides a valid estimate of the inverse susceptibility $(\mathrm{d}M/\mathrm{d}H)^{-1}$. We observe Curie-Weiss-like behavior above $T_{\mathrm{N}}$ (dark solid line) characterized by a fluctuating Curie-Weiss moment of $m_{\mathrm{CW}} = (3.95\pm0.05)~\mu_{\mathrm{B}}/\mathrm{f.u.}$ and a Curie-Weiss temperature of $\mathit{\Theta}_{\mathrm{CW}} = -(160\pm8)$~K. 

The fluctuating moment is distinctly smaller than the values reported in the literature~\cite{2006:Boeuf:PhysRevB, 1970:Endo:JPhysSocJpn, 1984:Helmholdt:JMagnMagnMater}. Note, however, that in samples incorporating magnetic impurity phases, such as Mn$_{2}$Sb, temperature sweeps of the susceptibility or $H/M$ typically exhibit a pronounced evolution as a function of field. The latter, in turn, may lead to an erroneous extrapolation of $m_{\mathrm{CW}}$ and $\mathit{\Theta}_{\mathrm{CW}}$. In our phase-pure specimens, the excellent agreement of the size of the fluctuating moment and the ordered moment, inferred from the neutron scattering experiments presented below and the literature~\cite{1968:Forster:JPhysChemSolids}, strongly implies a local-moment character of the magnetism in CuMnSb. Furthermore, the ratio $f = -\mathit{\Theta}_{\mathrm{CW}} / T_{\mathrm{N}} \approx 3$ suggests that geometric frustration may play a role~\cite{1994:Ramirez:AnnuRevMaterSci}.

\begin{figure}
\includegraphics[width=1.0\linewidth]{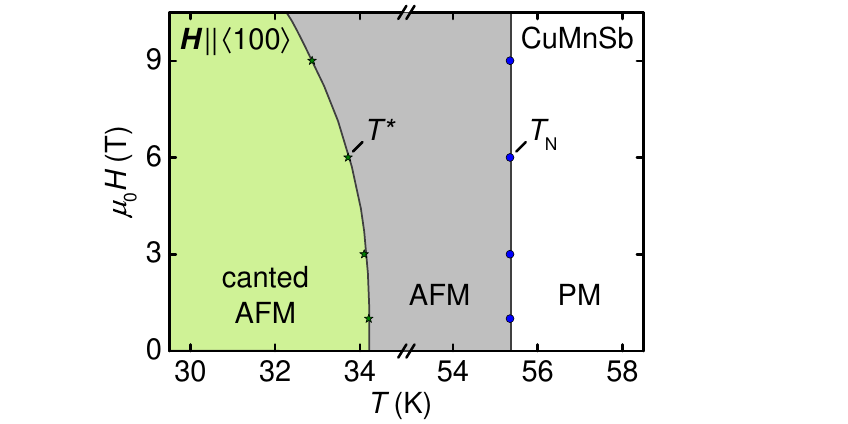}
\caption{Magnetic phase diagram of CuMnSb. The temperatures $T^{*}$ and $T_{\mathrm{N}}$ are inferred from the temperature derivative of the normalized magnetization, $\mathrm{d}M/H\mathrm{d}T$. We distinguish three regimes; paramagnet (PM), antiferromagnet (AFM), and canted antiferromagnet.}
\label{figure05}
\end{figure}

The N\'{e}el temperature $T_{\mathrm{N}}$ and the temperature of the cusp at $T^{*}$ are tracked best in the temperature derivative of the normalized magnetization, $\mathrm{d}M/H\mathrm{d}T$, depicted in Fig.~\ref{figure04}(c). Here, we associate $T_{\mathrm{N}}$ with a zero crossing and $T^{*}$ with a local maximum. The resulting magnetic phase diagram is shown in Fig.~\ref{figure05}. While $T_{\mathrm{N}}$ is independent of the magnetic field, $T^{*}$ is suppressed by a few Kelvin within the field range studied. Three regimes may be distinguished; a paramagnet (PM) at high temperatures, a commensurate antiferromagnet (AFM) at intermediate temperatures, and a canted antiferromagnet at low temperatures.

\begin{figure}
\includegraphics[width=1.0\linewidth]{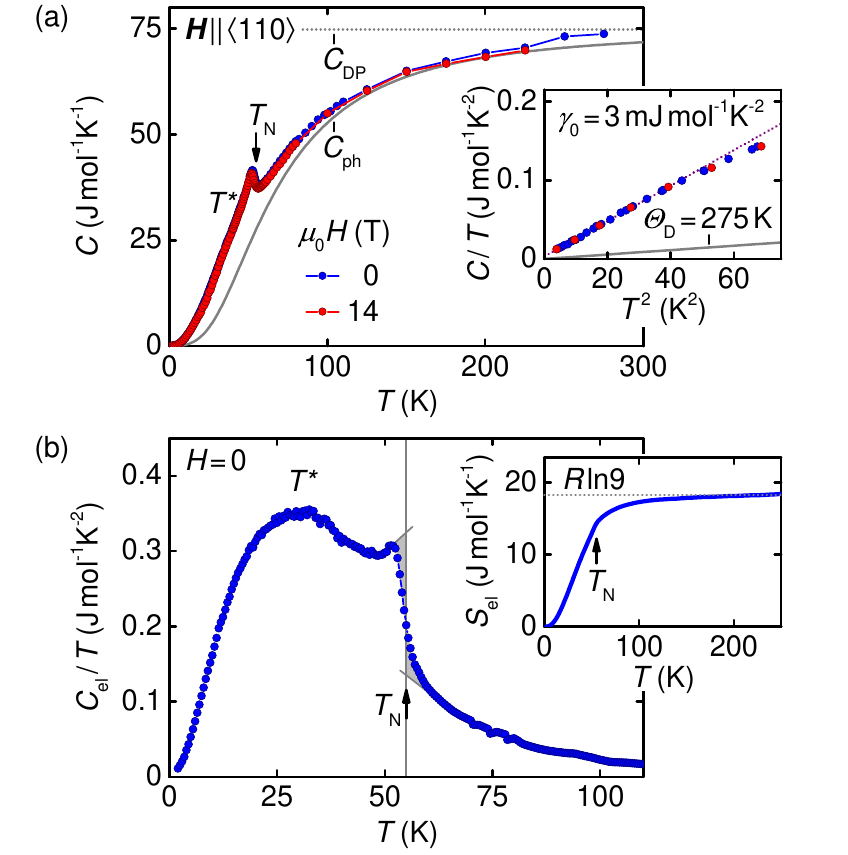}
\caption{Specific heat and entropy of CuMnSb. (a)~Specific heat as a function temperature. A clear lambda anomaly is observed at the N\'{e}el transition. The phonon contribution $C_{\mathrm{ph}}$ may be approximated by a Debye model with the Debye temperature $\mathit{\Theta}_{\mathrm{D}} = 275$~K. Magnetic fields up to 14~T show no substantial effect. Inset: Determination of the Sommerfeld coefficient $\gamma_{0}$. (b)~Specific heat after subtraction of the phonon contribution divided by temperature, $C_{\mathrm{el}}/T$, as a function of temperature. The N\'{e}el temperature is inferred by means of a entropy-conserving construction. Around $T^{*}$ a broad maximum is observed. Inset: Non-phonon contribution to the entropy $S_{\mathrm{el}}$.}
\label{figure06}
\end{figure}

These findings are corroborated by the specific heat shown in Fig.~\ref{figure06}(a). At high temperatures, the specific heat approaches the Dulong-Petit limit of $C_{\mathrm{DP}} = 9R$ and is dominated by phonon contributions. The latter may be approximated very well in terms of a Debye model with a Debye temperature $\mathit{\Theta}_{\mathrm{D}} = 275$~K (solid gray line). As illustrated in the inset, when the phonons freeze out at low temperatures, we extract a Sommerfeld coefficient $\gamma_{0} = 3~\mathrm{mJ\,mol}^{-1}\mathrm{K}^{-2}$. This small value is characteristic of a material with only weak electronic correlations and hence a local-moment nature of the magnetism. At the N\'{e}el temperature $T_{\mathrm{N}}$, a clear lambda anomaly is observed, indicating a second-order phase transition. A faint bulge may be perceived around $T^{*}$. Magnetic fields up to 14~T possess no significant influence on the specific heat curve.

For further analysis, we subtract the phonon contribution from the measured data. The remaining contribution to the specific heat divided by temperature, $C_{\mathrm{el}}/T$, is depicted in Fig.~\ref{figure06}(b). The value of $T_{\mathrm{N}}$ as inferred from a entropy-conserving construction (gray shading) is in excellent agreement with the value of $T_{\mathrm{N}}$ inferred from the magnetization. Around $T^{*}$ a broad maximum provides evidence of considerable contributions to the specific heat associated with canting of the magnetic structure. Numerically integrating $C_{\mathrm{el}}/T$ yields the associated entropy $S_{\mathrm{el}}$ as shown in the inset of Fig.~\ref{figure06}(b). Around $T_{\mathrm{N}}$ most of the entropy has been released and the slope distinctly changes. At high temperatures, $S_{\mathrm{el}}$ approaches a value of $R\,\ln9$, consistent with a local moment of $4~\mu_{\mathrm{B}}$.


\subsection{Electrical and thermal transport properties}
\label{Transport}

\begin{figure}
\includegraphics[width=1.0\linewidth]{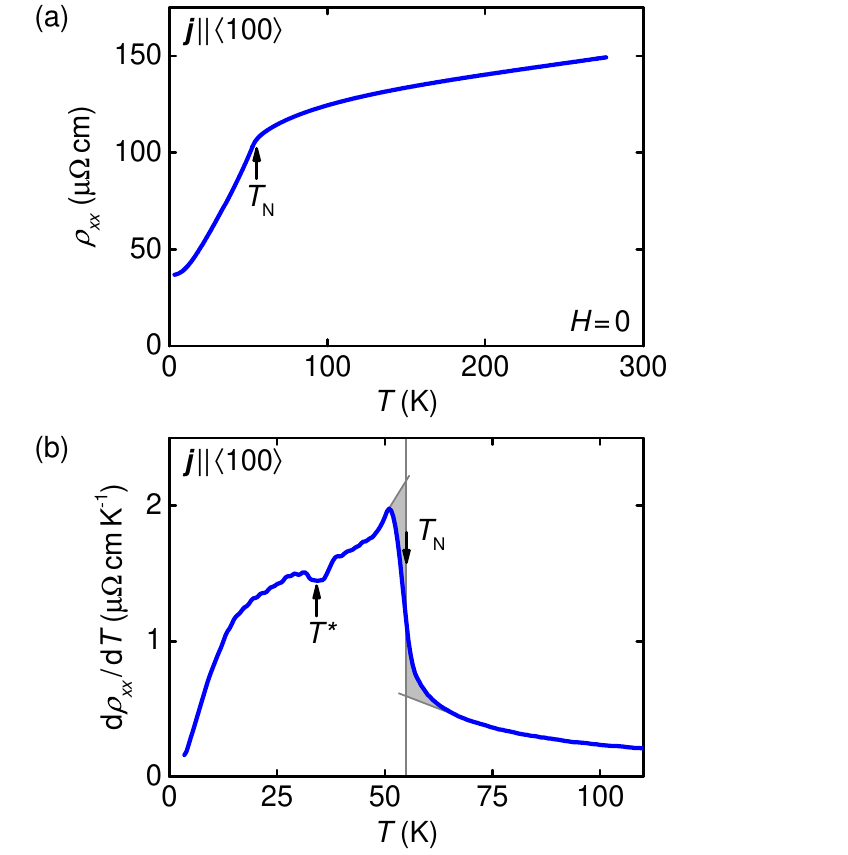}
\caption{Electrical resistivity of CuMnSb. (a)~Electrical resistivity as a function of temperature for current along $\langle100\rangle$. A kink marks the N\'{e}el temperature $T_{\mathrm{N}}$. Inset: Temperature dependence of the exponent $\alpha$ of a polynomial description of the measured data. (b)~Temperature derivative of the electrical resistivity, $\mathrm{d}\rho_{xx}/\mathrm{d}T$, as a function of temperature. The N\'{e}el temperature is inferred in analogy to the entropy-conserving construction in specific heat data. Around $T^{*}$ a minimum is observed.}
\label{figure07}
\end{figure}

Shown in Fig.~\ref{figure07}(a) is the electrical resistivity, $\rho_{xx}$, of single-crystal CuMnSb as a function of temperature for current along $\langle100\rangle$. As a function of decreasing temperature the electrical resistivity decreases monotonically. A distinct kink is associated with the onset of antiferromagnetic order at $T_{\mathrm{N}}$. We infer a residual resistivity of $\rho_{0} = 37~\mu\Omega\,\mathrm{cm}$, which is slightly smaller than the values reported in the literature~\cite{1982:Schreiner:SolidStateCommun, 1989:Otto:JPhysCondensMatter, 1995:Kirillova:PhysStatusSolidiB, 2006:Boeuf:PhysRevB, 2012:Maca:JMagnMagnMater}. The residual resistivity ratio is 4.2. While the value of $\rho_{0}$ is comparatively high for a transition metal compound, it is rather typical for Heusler compounds with their inherent proneness to structural disorder.  

Similar to the magnetization, the detailed position of $T_{\mathrm{N}}$ and $T^{*}$ may be inferred most accurately from the temperature derivative of the resistivity depicted in Fig.~\ref{figure07}(b). Note that the general shape of $\mathrm{d}\rho_{xx}/\mathrm{d}T$ strongly resembles $C_{\mathrm{el}}/T$. This suggests that the scattering observed in the electrical resistivity follows Fermi's golden rule with the corresponding density of state dominating the specific heat. The N\'{e}el temperature $T_{\mathrm{N}}$ is defined in analogy to the entropy-conserving construction of the specific heat. In comparison, in the vicinity of $T^{*}$ $\mathrm{d}\rho_{xx}/\mathrm{d}T$ displays a local minimum, while a broad maximum is observed in $C_{\mathrm{el}}/T$ indicating that this anomaly may be associated with a different type of scattering mechanism. The values of both $T_{\mathrm{N}}$ and $T^{*}$ are in excellent agreement with the values extracted from other quantities.

\begin{figure}
\includegraphics[width=1.0\linewidth]{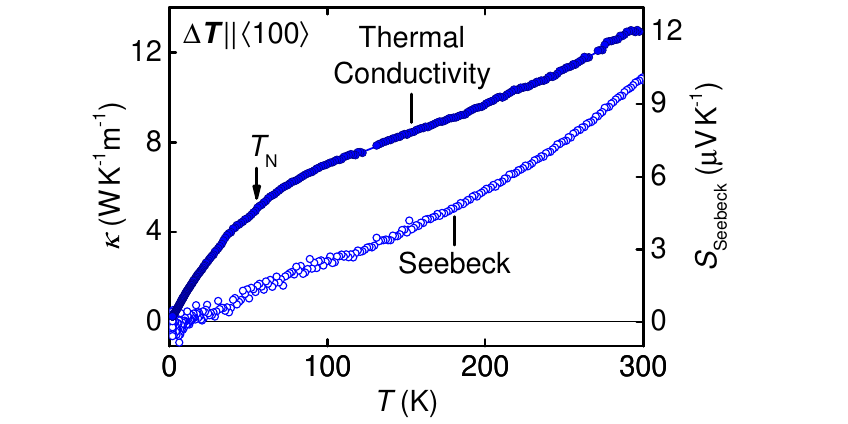}
\caption{Thermal transport properties of CuMnSb. Thermal conductivity (solid symbols) and Seebeck coefficient (open symbols) as a function of temperature for a temperature gradient along $\langle100\rangle$. A weak kink marks the N\'{e}el temperature $T_{\mathrm{N}}$ in the thermal conductivity.}
\label{figure08}
\end{figure}

In contrast to the electrical resistivity, the signatures of the magnetic ordering transitions are less pronounced in the thermal transport properties of CuMnSb. As shown in Fig.~\ref{figure08}, the thermal conductivity $\kappa$ decreases monotonically as a function of decreasing temperatures. Both the general shape of the curve and the absolute value of the conductivity are characteristic of a bad metal. Around $T_{\mathrm{N}}$ a weak change of slope is observed. The Seebeck coefficient $S_{\mathrm{Seebeck}}$ is positive and decreases monotonically as a function of decreasing temperatures, exhibiting no anomalies at $T_{\mathrm{N}}$ or $T^{*}$.

\begin{figure}
\includegraphics[width=1.0\linewidth]{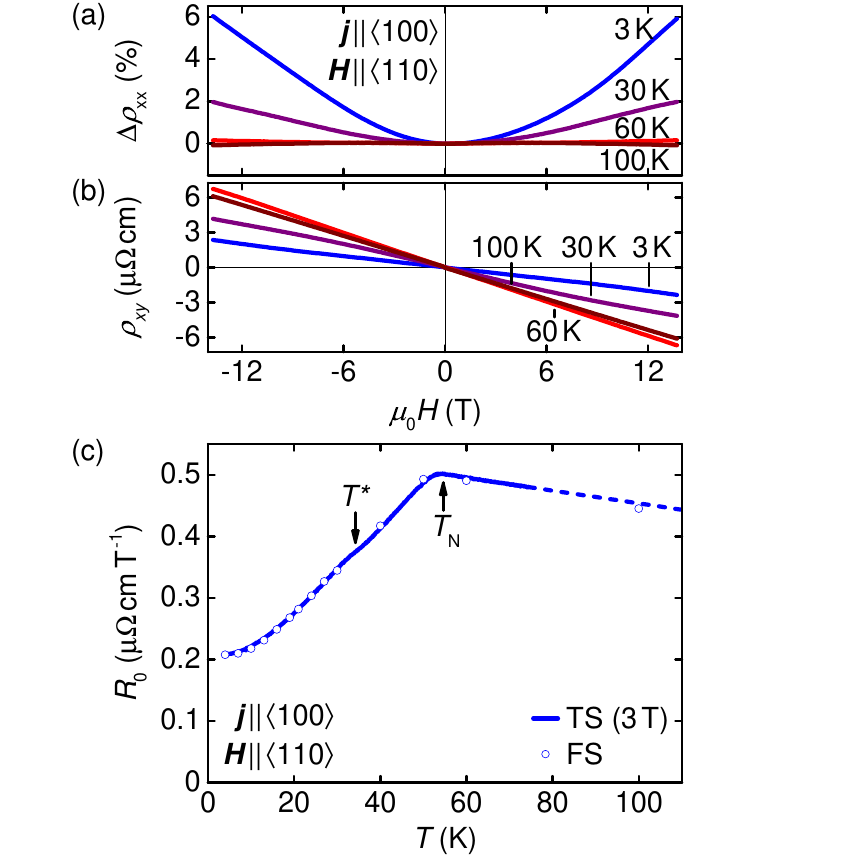}
\caption{Magnetoresistance and Hall effect of CuMnSb. (a)~Relative change of the electrical resistivity as a function of field for typical temperatures. (b)~Field dependence of the Hall effect for typical temperatures. (c)~Temperature dependence of the normal Hall coefficient $R_{0}$. The solid line is determined from temperature sweeps at $\pm3$~T, the open symbols are inferred from field sweeps.}
\label{figure09}
\end{figure}

Figure~\ref{figure09}(a) shows the magnetoresistance, $\Delta\rho_{xx} = (\rho_{xx}/\rho_{xx}^{H=0}) - 1$. Under increasing field the resistance increases quadratically at low temperatures, reaching about $+6\%$ at 14~T. With increasing temperature the magnetoresistance decreases and turns very weakly negative between 60~K and 100~K with very little curvature.

The Hall resistivity $\rho_{xy}$ is essentially linear as a function of field as shown in Fig.~\ref{figure09}(b). The slope is negative, resulting in a negative Hall constant $-R_{0}$ (we note that the minus sign arises from the definition of the Hall constant in the Hall conductivity). The temperature dependence of $R_{0}$ is best extracted from temperature sweeps of the Hall resistivity in fixed fields via $R_{0} = -\rho_{xy}/(\mu_{0}H)$. Corresponding data inferred from measurements at $\pm3$~T are shown as solid line in Fig.~\ref{figure09}(c). Open symbols denote values derived from field sweeps. The dashed line is a guide to the eye. The N\'{e}el temperature $T_{\mathrm{N}}$ is associated with a maximum and a distinct kink in the Hall constant. At $T^{*}$ we observe a weak change of slope. From $R_{0}$ we estimate an averaged charge carrier concentration $n = (R_{0}e)^{-1}$ of the order of $2\cdot10^{21}~\mathrm{cm}^{-3}$. This value is typical for a bad metal, suggesting a small density of states at the Fermi level, and the positive sign indicates dominant hole-like conduction. Our results are consistent with the calculations of Jeong \textit{et al.} predicting a semi-metallic state with heavy electron masses and normal hole masses~\cite{2005:Jeong:PhysRevB}.


\subsection{Powder neutron diffraction}
\label{Powder}

\begin{figure*}
\includegraphics[width=1.0\linewidth]{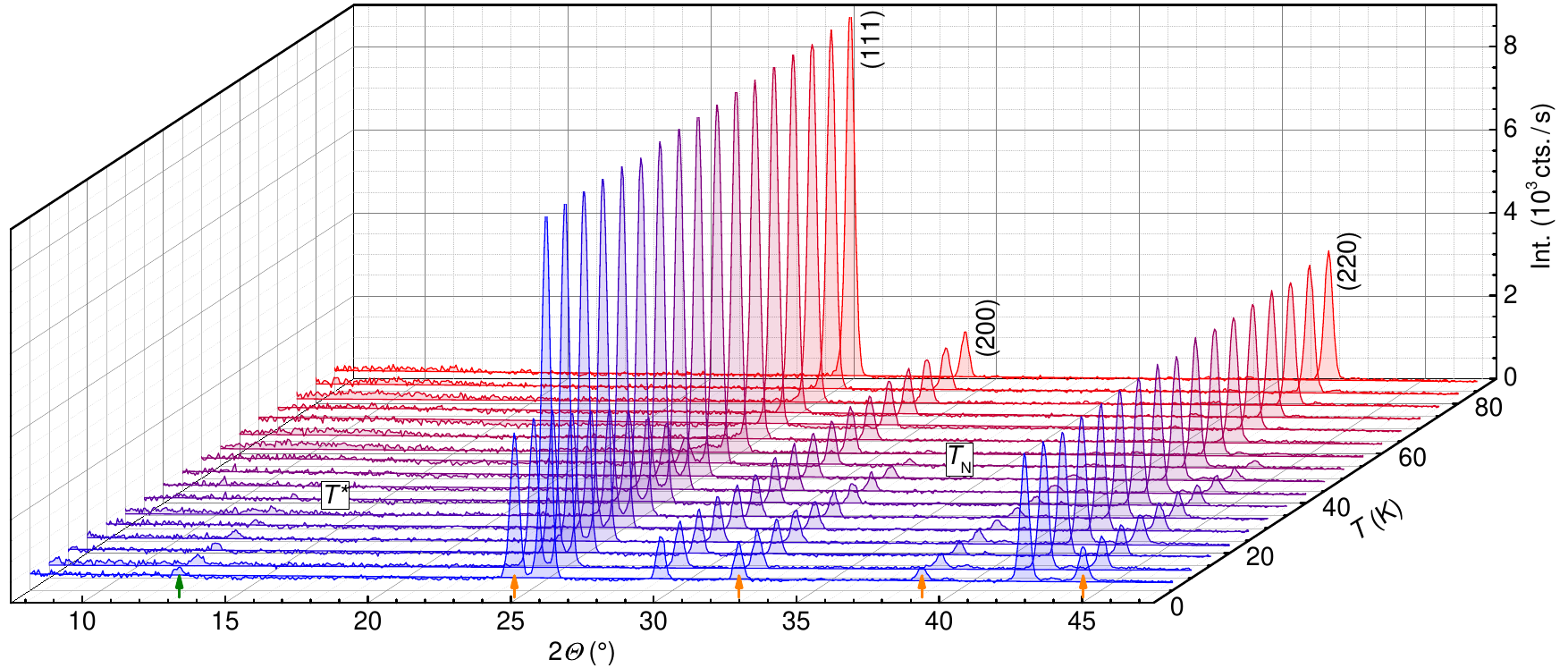}
\caption{Powder neutron diffraction as recorded at SPODI. Diffraction data are shown for temperatures between 85~K (red curve) and 4~K (blue curve). In addition to the intensity maxima due to nuclear scattering, a set of magnetic intensity maxima emerges with decreasing temperature below $T_{\mathrm{N}}$ (orange arrows), followed by one additional peak at small angles below $T^{*}$ (green arrow at $2\mathit{\Theta} = 12.7^{\circ}$).}
\label{figure10}
\end{figure*}

Typical data observed in powder neutron diffraction at SPODI for temperatures between 4~K and 85~K and small scattering angles are shown in Fig.~\ref{figure10}. At temperatures above the N\'{e}el temperature $T_{\mathrm{N}}$ (red curves), Bragg peaks corresponding to the nuclear scattering according to space group $F\bar{4}3m$ only are observed, in excellent agreement with the X-ray diffraction data, cf.\ Fig.~\ref{figure03}(a). Below $T_{\mathrm{N}}$ (purple curves), a set of additional maxima emerges (marked by orange arrows) that increase with decreasing temperature, characteristic of magnetic order. Below $T^{*}$ (blue curves), an additional intensity maximum appears at a small angle, $2\mathit{\Theta} = 12.7^{\circ}$. Its intensity is rather weak but clearly discernible, increasing as a function of decreasing temperature.

In order to account for the observed diffraction patterns, we superimpose the nuclear structure in space group $F\bar{4}3m$ with a magnetic structure. For these refinements, we attribute the magnetic moment exclusively to the Mn sites and use the magnetic form factor of the Mn$^{2+}$ ion, according to the description of CuMnSb as self-doped Cu$^{1+}$Mn$^{2+}$Sb$^{3-}$~\cite{2005:Jeong:PhysRevB}. Assuming that the magnetic propagation vector is of the form $\bm{k} = \frac{1}{2}\langle111\rangle$, consistent with previous reports~\cite{1968:Forster:JPhysChemSolids}, we find excellent agreement with our experimental data. Note that there are four independent domains $\bm{k}_{i}$ as described by $\bm{k}_{1} = \frac{1}{2}[111]$, $\bm{k}_{2} = \frac{1}{2}[\bar{1}\bar{1}1]$, $\bm{k}_{3} = \frac{1}{2}[\bar{1}1\bar{1}]$, and $\bm{k}_{4} = \frac{1}{2}[1\bar{1}\bar{1}]$. 

For the half-Heusler environment of CuMnSb, the irreducible representations leaving $\bm{k}$ invariant yield five non-centrosymmetric magnetic space groups. Group $R[I]3m$ is non-magnetic and may be ignored, while fits using group $P[I]1$ did not converge. The three remaining groups are $R[I]3c$, $C[B]m$, and $C[B]c$. In our cubic setting, $R[I]3c$ corresponds to moments that are collinear to $\bm{k}$, i.e., oriented along $\langle111\rangle$. In $C[B]m$, the moments point along a $\langle110\rangle$ axes perpendicular to $\bm{k}$. In $C[B]c$, the moments may possess finite components both parallel and perpendicular to the propagation vector. 

\begin{figure}
\includegraphics[width=1.0\linewidth]{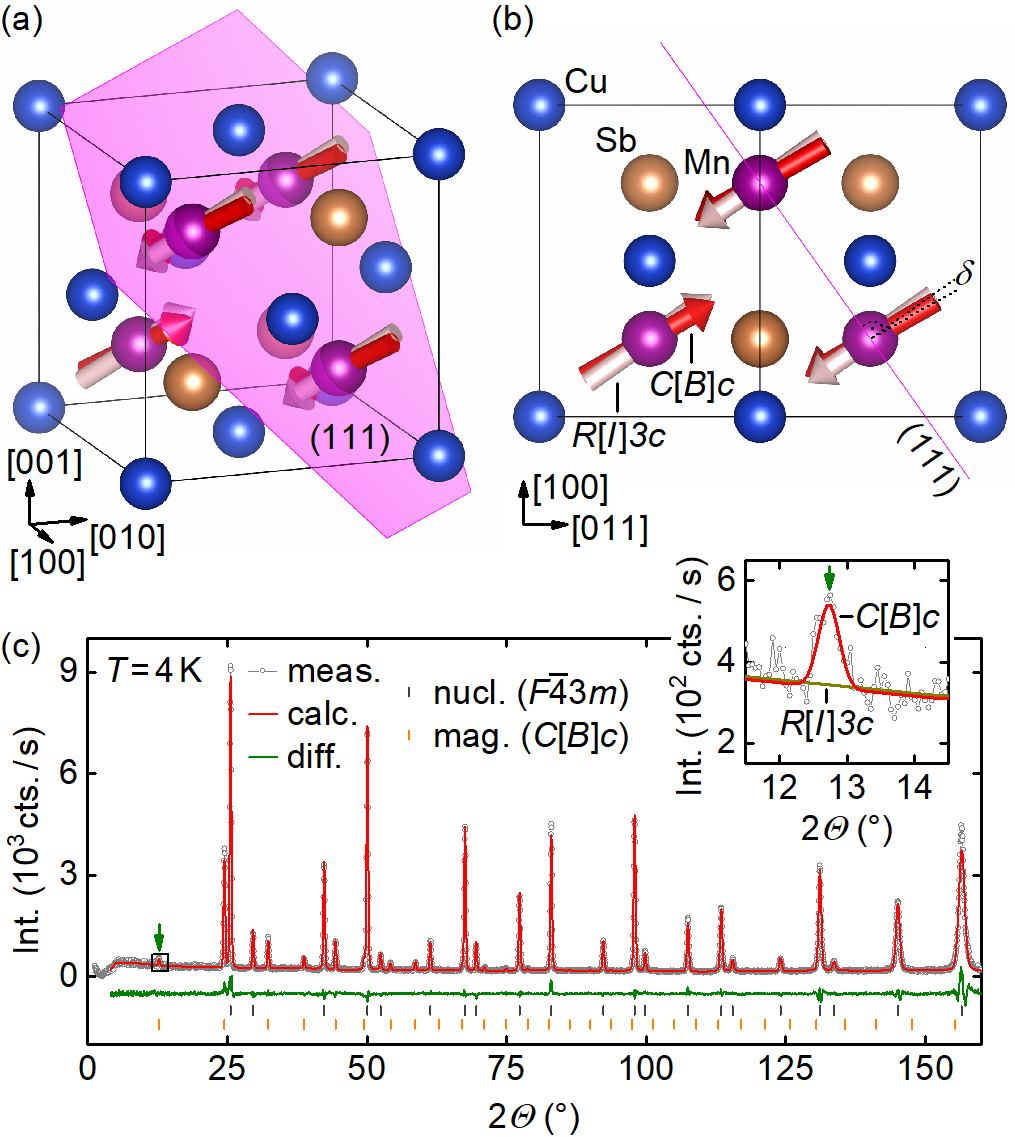}
\caption{Refinement of powder neutron diffraction data of CuMnSb. \mbox{(a),(b)}~Schematic illustration of the relevant magnetic space groups. In $R[I]3c$ the moments are perpendicular to the nuclear $(111)$ plane (rose arrows), in $C[B]c$ they are canted by the angle $\delta$ (red arrows). (c)~Low-temperature diffraction data ($T = 4$~K) and Rietveld refinement based on the nuclear space group $F\bar{4}3m$ and the magnetic space group $C[B]c$. Inset: Enlarged view illustrating the difference between the magnetic space groups $R[I]3c$ and $C[B]c$.}
\label{figure11}
\end{figure}

In Ref.~\onlinecite{1968:Forster:JPhysChemSolids}, commensurate type-II antiferromagnetic order was suggested comprising magnetic moments parallel to the $\langle111\rangle$ axes. These moments are ferromagnetically aligned within the respective $\{111\}$ plane, while neighboring planes couple antiferromagnetically with each other. This spin arrangement is consistent with space group $R[I]3c$ and schematically illustrated by the rose arrows in Figs.~\ref{figure11}(a) and \ref{figure11}(b). We find that for temperatures $T^{*} < T < T_{\mathrm{N}}$ Rietveld refinements of our diffraction data are in excellent agreement with this magnetic structure. In particular, the lack of a magnetic $\frac{1}{2}(111)$ satellite at $2\mathit{\Theta} = 12.7^{\circ}$, cf.\ Fig.~\ref{figure10}, implies that the magnetic moments are aligned parallel to $\langle111\rangle$. 

In turn, the emergence of the latter maximum for temperatures $T < T^{*}$ indicates that a canting away from the $\langle111\rangle$ direction occurs, described by the angle $\delta$. The resulting magnetic space group is $C[B]c$, where $R[I]3c$ corresponds to $C[B]c$ with $\delta = 0$. Neighboring moments cant in opposite directions as sketched by the red arrows in Figs.~\ref{figure11}(a) and \ref{figure11}(b). Consequently, no net ferrimagnetic moment is expected, consistent with the magnetization data. As shown in Fig.~\ref{figure11}(c), a Rietveld refinement based on this canted antiferromagnetic structure is in excellent agreement with the experimental low-temperature diffraction data up to high diffraction angles. The inset highlights the salient difference of the refinements for the magnetic space groups $R[I]3c$ and $C[B]c$, i.e., that the maximum at $12.7^{\circ}$ may only be explained by the latter. Refinements using the magnetic space group $C[B]m$ are significantly worse than those using $R[I]3c$ or $C[B]c$ (not shown).

\begin{figure}
\includegraphics[width=1.0\linewidth]{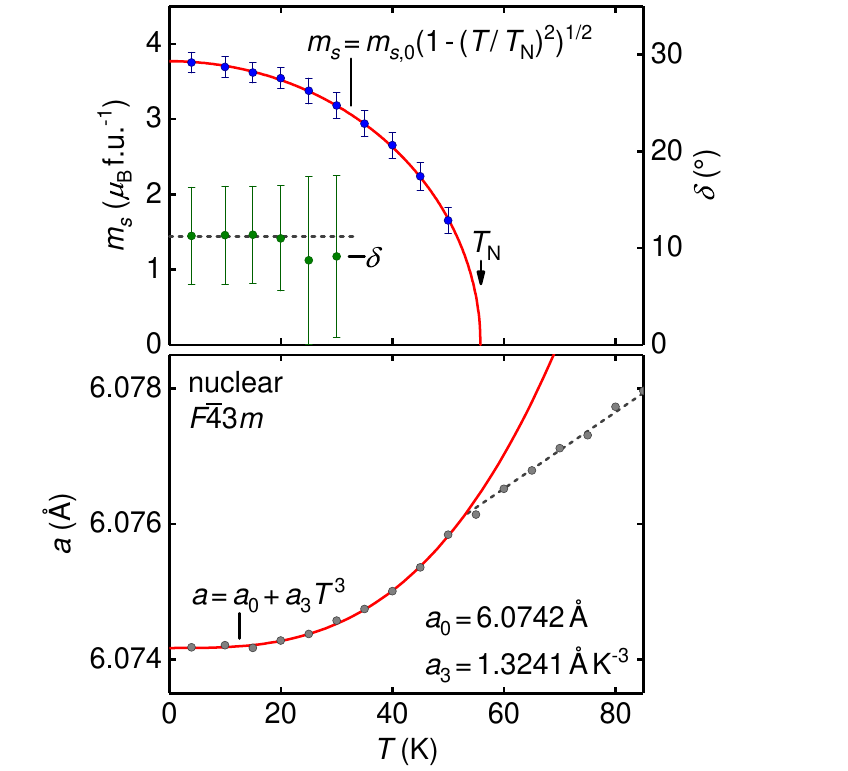}
\caption{Ordered magnetic moment, tilting angle, and lattice parameter inferred from the refinement of the powder neutron data. (a)~Temperature dependence of the magnetic moment $m_{s}$ and the angle $\delta$. (b)~Lattice constant $a$ as a function of temperature. The solid red lines are fits to the data. The dashed lines represent guides to the eye.}
\label{figure12}
\end{figure}

Figure~\ref{figure12}(a) shows the temperature dependence of the ordered magnetic moment $m_{s}$ as inferred from the refinements described above. We extract a value of 3.8~$\mu_{\mathrm{B}}/\mathrm{f.u.}$ at low temperatures, in good agreement with both our magnetization data and previous neutron scattering studies~\cite{1968:Forster:JPhysChemSolids}. As a function of increasing temperature, the moment decreases according to $m_{s} / m_{s,0} = \sqrt{1 - (T/T_{\mathrm{N}})^{2}}$ and vanishes at $T_{\mathrm{N}}$, as indicated by the red solid line. Within the resolution of the present data, the canting angle $\delta$ remains at about $11^{\circ}$ for $T < T^{*}$. 

The temperature dependence of the nuclear lattice constant inferred from the same refinements is depicted in Fig.~\ref{figure12}(b). The absolute values are in good agreement with the results of our X-ray diffraction. Moreover, due to the finer temperature steps recorded in neutron scattering, we are able to resolve a weak change of slope around $T_{\mathrm{N}}$. We attribute this finding to magnetostriction, where the lattice constant increases with the third power of the temperature for $T < T_{\mathrm{N}}$ while exhibiting an essentially linear behavior for larger temperatures.


\subsection{Single-crystal neutron diffraction}
\label{SingleCrystal}

\begin{figure}
\includegraphics[width=1.0\linewidth]{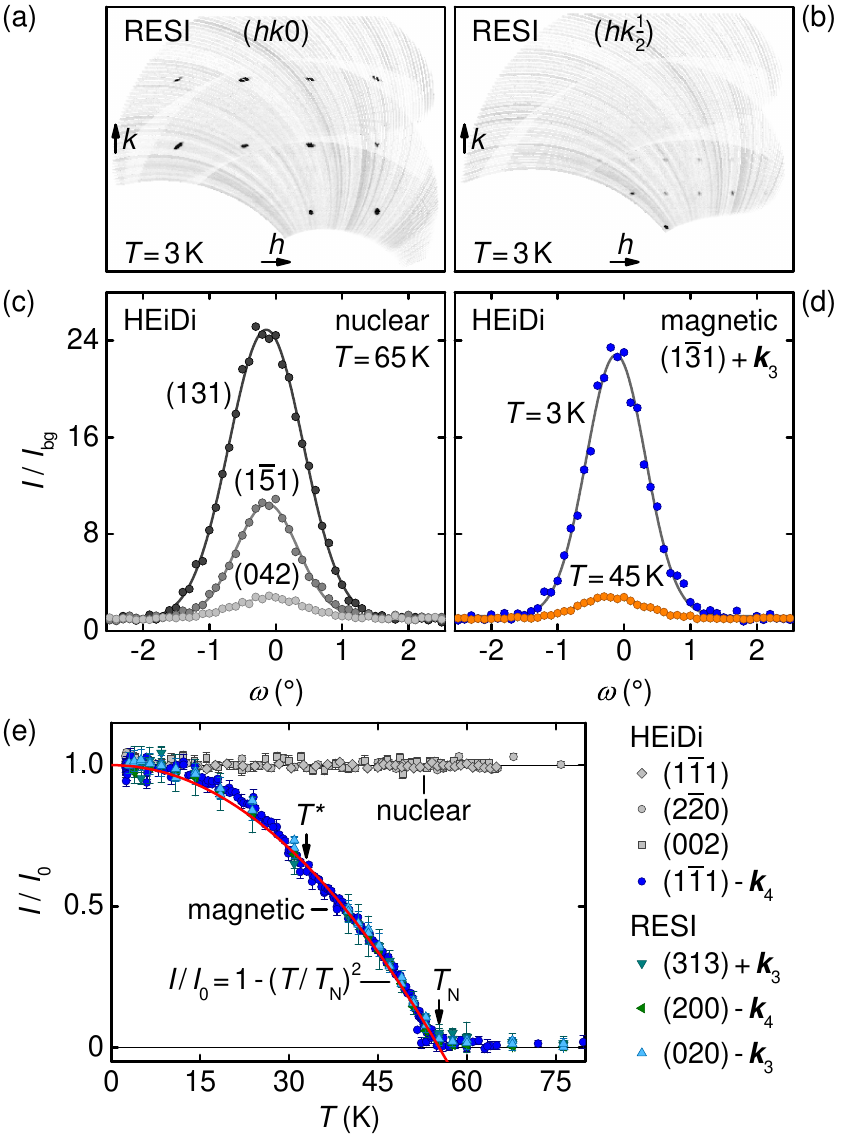}
\caption{Single-crystal neutron diffraction data recorded at RESI and HEiDi. (a)~Intensity distribution in the $(hk0)$ plane at low temperature. Maxima (dark contrast) arise from the nuclear structure. (b)~Intensity distribution in the $(hk\frac{1}{2})$ plane. Maxima at half reciprocal lattice spacing are attributed to the commensurate antiferromagnetic structure. (c)~Rocking scans for typical nuclear Bragg peaks. Gaussian fits (solid lines) indicate a full width at half maximum of about $1^{\circ}$. (d)~Rocking scans for a typical magnetic reflection at two temperatures. (e)~Normalized intensity for typical nuclear (gray symbols) and magnetic (colored symbols) Bragg peaks as a function of temperature. The red solid red line is a fit to the magnetic data. At $T^{*}$ a weak change of slope is observed.}
\label{figure13}
\end{figure}

The magnetic structure proposed above as based on the refinement of the powder diffraction data is corroborated by single-crystal neutron diffraction data. Shown in Figs.~\ref{figure13}(a) and \ref{figure13}(b) are maps of the the reciprocal space planes $(hk0)$ and $(hk\frac{1}{2})$, recorded on the diffractometer RESI at low temperatures. Maxima in the $(hk0)$ plane are attributed to the half-Heusler nuclear structure. In contrast, maxima in the $(hk\frac{1}{2})$ plane are characteristic of the commensurate antiferromagnetic order with a doubling of the magnetic unit cell in real space. 

Rocking scans of a large number of nuclear and magnetic Bragg peaks were carried out on the diffractometer HEiDi. Typical data for three different reflections are shown in Fig.~\ref{figure13}(c). The maxima are symmetric and very well described by Gaussian fits (solid lines). We obtain full widths at half maximum of about $1^{\circ}$, indicating a small mosaicity for this type of compound and in turn an excellent sample quality. Rocking scans on magnetic reflections, illustrated for $(1\bar{3}1) + \bm{k}_{3}$ in Fig.~\ref{figure13}(d), may also be described by Gaussians of similar full width at half maximum. In contrast to nuclear peaks, however, their intensity strongly depends on temperature as presented in Fig.~\ref{figure13}(e).

In order to compare the different maxima, their intensity has been normalized by the intensity inferred from an extrapolation to zero temperature. While the nuclear Bragg peaks are essentially independent of temperature in the temperature range studied, the magnetic intensity vanishes at the N\'{e}el temperature $T_{\mathrm{N}}$. As indicated by the red solid line, with increasing temperature the intensity decreases as $I / I_{0} = 1 - (T/T_{\mathrm{N}})^{2}$, i.e., it scales with the square of the ordered magnetic moment. At $T^{*}$, a distinct change of slope is observed for all magnetic reflections.

\begin{figure}
\includegraphics[width=1.0\linewidth]{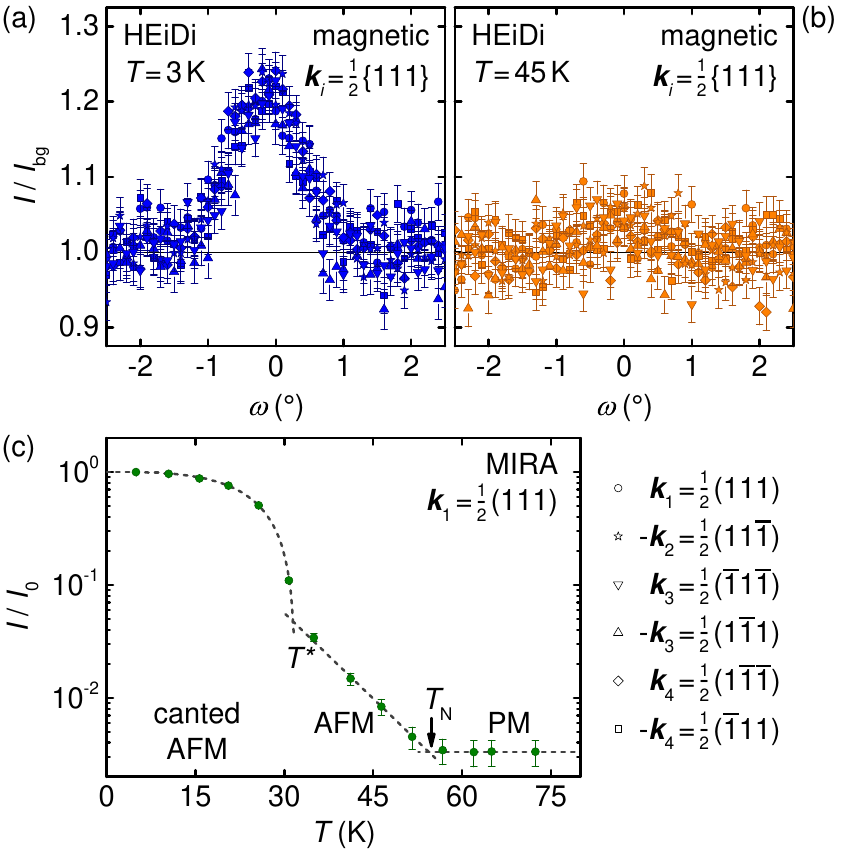}
\caption{Single-crystal neutron diffraction data of the canting of the antiferromagnetic order as recorded at HEiDi and MIRA. \mbox{(a),(b)}~Rocking scans around the $\frac{1}{2}\{111\}$ positions characteristic of the canted antiferromagnetism in the magnetic space group $C[b]c$ at temperatures well below $T^{*}$ and $T^{*} < T < T_{\mathrm{N}}$, respectively. (c)~Temperature dependence of the intensity of the $\frac{1}{2}(111)$ maximum on a logarithmic scale. The dashed lines are guides to the eye.}
\label{figure14}
\end{figure}

The change of slope is a consequence of the spin canting at low temperatures, $T < T^{*}$. The most prominent hallmark of the latter, however, is the emergence of weak magnetic intensity maxima at the $\frac{1}{2}\{111\}$ positions illustrated in Fig.~\ref{figure14}. For the cubic half-Heusler structure four antiferromagnetic domains may be expected, denoted $\bm{k}_{i}$, with a total of eight magnetic satellites at $\pm\bm{k}_{1}$, $\pm\bm{k}_{2}$, $\pm\bm{k}_{3}$, and $\pm\bm{k}_{4}$. We carefully checked the corresponding positions in reciprocal space on HEiDi and observed all six satellites that were accessible experimentally. As shown in Fig.~\ref{figure14}(a), the maxima may be described by Gaussians of a full width at half maximum of about $1^{\circ}$, i.e., akin to all other Bragg peaks. The observation of essentially identical intensities of all satellites indicates that the four antiferromagnetic domains are populated equally.

Refinements of the data recorded at HEiDi at low temperature, taking into account a large number of magnetic Bragg peaks (not shown), are also in excellent agreement with a magnetic structure in space group $C[B]c$. We obtain ordered moments of 3.8~$\mu_{\mathrm{B}}/\mathrm{f.u.}$ and a canting angle $\delta = 14^{\circ}$, both perfectly consistent with our powder neutron data within the error bars. As shown in Fig.~\ref{figure14}(b), there are no $\frac{1}{2}\{111\}$ satellites at temperatures between $T^{*}$ and $T_{\mathrm{N}}$, consistent with zero canting angle or the magnetic space group $R[I]3c$, respectively. 

The detailed temperature dependence of the $\frac{1}{2}(111)$ intensity is finally depicted in Fig.~\ref{figure14}(c). These data were recorded at the spectrometer MIRA in triple-axis geometry by integrating the intensity around zero energy transfer. Note the logarithmic intensity scale. As a function of increasing temperature the intensity monotonically decreases, where three regimes may be distinguished. At low temperatures, the satellite intensity decreases $I \propto T^{3.2}$, essentially vanishing at $T^{*}$. At intermediate temperatures $T^{*} < T < T_{\mathrm{N}}$ small but finite intensity is observed, putatively indicating spin fluctuations that preempt the spin canting at lower temperatures. For high temperatures $T > T_{\mathrm{N}}$, CuMnSb is paramagnetic and neither at $\frac{1}{2}(111)$ nor at any other magnetic Bragg position significant intensity is observed. 

\begin{figure}
\includegraphics[width=1.0\linewidth]{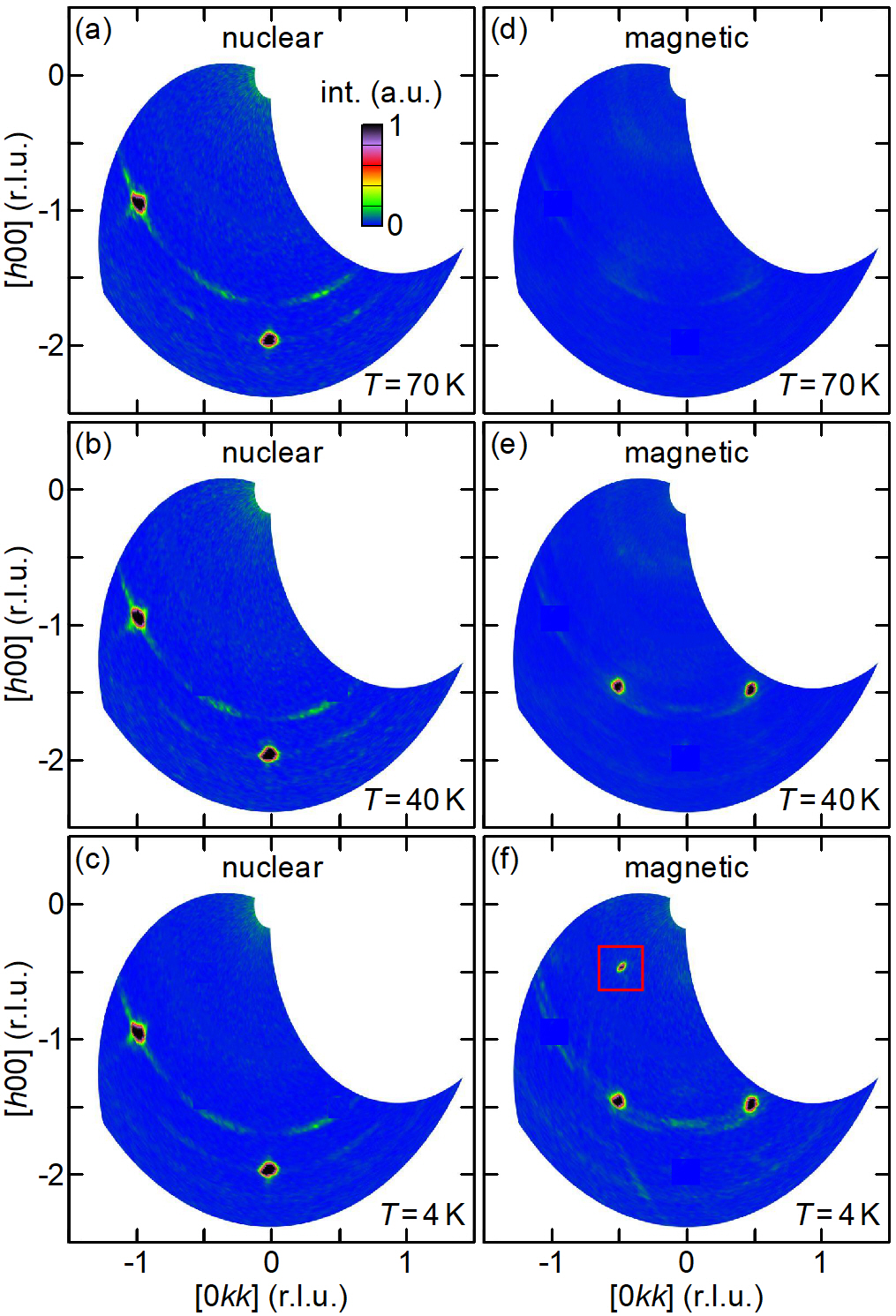}
\caption{Single-crystal neutron diffraction data of CuMnSb recorded at DNS. \mbox{(a)--(c)}~Intensity distribution of nuclear contributions in the $(hkk)$ plane for three different temperatures. (d)~Magnetic intensity distribution for $T > T_{\mathrm{N}}$. No maxima are observed. (e)~For $T < T_{\mathrm{N}}$ maxima appear at positions consistent with commensurate antiferromagnetic order. (f)~For $T < T^{*}$ an additional maximum at $\frac{1}{2}(\bar{1}\bar{1}\bar{1})$ is characteristic of finite spin canting (red square). Note that the peak intensities of the nuclear reflections correspond to ${\sim}10^{2}$ on the linear color scale. Remnants from the nuclear maxima are attributed to insufficient flipping ratio corrections and were manually masked.}
\label{figure15}
\end{figure}

Single-crystal neutron diffraction data as recorded at DNS, depicted in Fig.~\ref{figure15}, corroborate and summarize our findings nicely. The very weak diffuse rings of intensity may be attributed to powder-like contributions stemming from the surface of the large single crystal. As shown in the left column, the nuclear contribution exhibits intense peaks at $(hkl)$ positions consistent with the half-Heusler structure, namely $(\bar{1}\bar{1}\bar{1})$ and $(\bar{2}00)$. As a function of decreasing temperature (top to bottom), the scattering pattern does not change. 

In contrast, the magnetic contribution (right column) exhibits a distinct evolution as a function of temperature. Above $T_{\mathrm{N}}$, see Fig.~\ref{figure15}(d), no magnetic signal is observed. Below $T_{\mathrm{N}}$, see Fig.~\ref{figure15}(e), peaks emerge that are consistent with commensurate antiferromagnetic structure and $\bm{k} = \frac{1}{2}\langle111\rangle$. Note the characteristic doubling of the magnetic compared to the nuclear unit cell. Below $T^{*}$, see Fig.~\ref{figure15}(f), a weak additional maximum appears at the $\frac{1}{2}(\bar{1}\bar{1}\bar{1})$ position, marked by the red square. This maximum is the characteristic of finite components of the magnetic moment perpendicular to $\bm{k}$, i.e., the canting of the moment direction away from $\bm{k}$.


\section{Conclusions}
\label{Conclusions}

In summary, to the best of our knowledge, we have grown for the first time large single crystals of the half-Heusler compound CuMnSb. Using a tiny Sb excess in the starting composition, the single crystal grown is phase-pure. Magnetization, specific heat, electrical resistivity, and Hall effect measurements on these phase-pure specimens consistently suggest a local-moment character of the antiferromagnetism in a metallic environment. These thermodynamic and transport quantities clearly exhibit anomalies at the onset of magnetic order at the N\'{e}el temperature $T_{\mathrm{N}} = 55$~K, as well as a second anomaly at $T^{*} \approx 34$~K well below $T_{\mathrm{N}}$ that has not been reported before. 

Below the N\'{e}el temperature $T_{\mathrm{N}} = 55$~K, our neutron scattering data identify commensurate type-II antiferromagnetic order with propagation vectors and magnetic moments aligned along the $\langle111\rangle$ directions, corresponding to the magnetic space group $R[I]3c$. This form of magnetic order is consistent with the fcc structure of the Mn sublattice and the well-understood antiferromagnetism in the transition metal oxides MnO, NiO, and CoO.

However, using powder and single-crystal neutron diffraction, we unambiguously connect the anomaly at $T^{*}$ with a spin canting, where the moments tilt away from $\langle111\rangle$ axes by a finite angle $\delta \approx 11^{\circ}$. Neither data recorded at RESI nor DNS suggest further diffraction peaks at other locations. Thus the scattering information appears to be complete. Taken together, a canted antiferromagnetic structure is stabilized without uniform magnetic moment, described by the magnetic space group $C[B]c$. This result appears to be rather surprising in view of the fcc Mn sublattice and the large ordered moments. 

Based on the fundamental symmetries of the crystal structure it is hard to reconcile the observed canting with subleading interactions. In contrast, our results are in excellent agreement with the calculations of the magnetic ground state by M\'{a}ca \textit{et al.}~\cite{2016:Maca:PhysRevB} and underscore that high-quality samples are crucial when trying to resolve the intrinsic properties of compounds that are sensitive to disorder and defects. Therefore, our findings will be of great interest not only for a wide range of Heusler compounds but also for other materials supporting magnetic order on fcc sublattices.


\begin{acknowledgments}
We wish to thank P.~B\"{o}ni, G.~Brandl, M.~Gangl, F.~Kortmann, J.~K\"{u}bler, D.~Mallinger, S.~Mayr, V.~Pet\v{r}\'{i}\v{c}ek, J.~Schilling, and A.~Schneidewind for fruitful discussions and assistance with the experiments. Parts of the data were collected on HEiDi, jointly operated by RWTH Aachen and Forschungszentrum J\"{u}lich GmbH (JARA collaboration). Financial support by the Deutsche Forschungsgemeinschaft (DFG) through TRR80 (projects E1 and F2) and by the European Research Council (ERC) through Advanced Grant 291079 (TOPFIT) is gratefully acknowledged. A.R.\ and A.B.\ acknowledge financial support through the TUM graduate school.
\end{acknowledgments}

\end{document}